\documentclass[pra,superscriptaddress,twocolumn,showpacs,floatfix]{revtex4}
\usepackage{bm}
\usepackage{graphicx} 

\begin{document}

\title{Quantum matter wave dynamics with moving mirrors}

\author{A. del Campo\footnote{E-mail: adolfo.delcampo@ehu.es}}
\author{J. G. Muga
}
\affiliation{Departamento de Qu\'\i mica-F\'\i sica,
UPV-EHU, Apartado 644, Bilbao, Spain}

\author{M. Kleber}
\affiliation{Physik-Department T30,
Technische Universit{\"a}t M{\"u}nchen, James-Franck Strasse, 85747 Garching, Germany}

\def\t{{\rm t}}
\def\K{{\rm K}}
\def\la{\langle}
\def\ra{\rangle}
\def\om{\omega}
\def\Om{\Omega}
\def\vep{\varepsilon}
\def\wh{\widehat}
\def\tr{\rm{Tr}}
\def\da{\dagger}
\def\iz{\left}
\def\zi{\right}
\newcommand{\beq}{\begin{equation}}
\newcommand{\eeq}{\end{equation}}
\newcommand{\beqa}{\begin{eqnarray}}
\newcommand{\eeqa}{\end{eqnarray}}
\newcommand{\intf}{\int_{-\infty}^\infty}
\newcommand{\into}{\int_0^\infty}

\begin{abstract}
When a stationary reflecting wall acting as a perfect mirror for an
atomic beam with well defined incident velocity is suddenly removed,
the density profile develops during the time evolution an
oscillatory pattern known as diffraction in time. The interference
fringes are suppressed or their visibility is diminished by several
effects such as averaging over a distribution of incident
velocities, apodization of the aperture function, atom-atom
interactions, imperfect reflection or environmental noise. However,
when the mirror moves with finite velocity along the direction of
propagation of the beam, the visibility of the fringes is enhanced.
For mirror velocities below beam velocity, as used for slowing down
the beam, the matter wave splits into three regions separated by
space-time points with classical analogues. For mirror velocities
above beam velocity a visibility enhancement occurs without a
classical counterpart. When the velocity of the beam approaches that
of the mirror the density oscillations rise by a factor 1.8 over the
stationary value.
\end{abstract}


\maketitle
\section{Introduction}

Coherent and intense cold atom beams stand up for their applications
in metrology, matter-wave interferometry or atom lithography
\cite{Berman97}. A fast beam can be decelerated down to cold or
ultracold temperatures by reflection from a mirror moving in the
direction of the atoms. An early example is the achievement of an
ultracold beam of neutrons by neutron reflection from a moving
Ni-surface [2]. Moving mirrors for manipulating cold atom waves have
been also produced with a time-modulated, blue-detuned evanescent
light wave propagating along the surface of a glass prism
\cite{Dalibard1,Dalibard2}. More recently bright beams of Helium
atoms have been successfully slowed down using a moving Si-crystal
on a spinning rotor \cite{Raizen06}; and Rb atoms with a moving
magnetic mirror on a conveyor belt \cite{Guery06}, which provides a
promising mechanism to generate a continuous, intense and very slow
beam of guided atoms.

\begin{figure}
\includegraphics[width=5cm,angle=0]{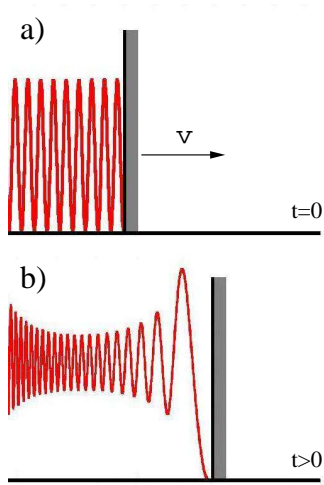}
  \caption{Setup for a matter-wave beam  reflected from a mirror moving with velocity $v$.
a) At time equal to zero the beam forms a standing matter wave. b)
During the time evolution, the density profile of the beam exhibits
a self-modulation enhanced with respect to the sudden removal of the
mirror, known as diffraction in time ($v=\infty$). }  \label{setup}
\end{figure}

The motivation of the present work is that, whereas the analysis of
the dynamics in such devices is usually based on classical
mechanics, the long de Broglie wavelengths in the ultracold regime
requires a quantum treatment. When operating in the quantum domain,
such mechanisms are expected to exhibit a more subtle dynamics than
simple classical trajectory reflection, with wave aspects becoming
more prominent. Therefore our aim here is to study a simple solvable
case which can be a reference for more complex settings. In this
task we shall be guided by the abundant work on the ``Moshinsky
shutter'': One of the most relevant quantum transient effects in
matter-wave beams is the diffraction in time effect
\cite{Moshinsky52,Moshinsky76,Kleber94,MMS99,DMM07}, 
an oscillatory self-modulation of the density profile of a
suddenly released beam. To date, it has been observed in a wide
variety of systems such as neutrons \cite{HFGGGW98}, ultracold atoms
\cite{atomdit}, electrons \cite{Lindner05} and even Bose-Einstein
condensates using vibrating mirrors \cite{CMPL05}. However, the
effect weakens with the width of the beam velocity distribution
\cite{DM05,MB00}, dissipation \cite{MS01}, environmental noise
\cite{DMM07}, finite-size of the beam and confinement
\cite{GK76,Godoy02,Godoy03,GM07}, strong interatomic interactions
\cite{DM06}, and the smoothing of the aperture function of the
shutter \cite{SM88,DM05,GM06}. Indeed, any variant discussed with
respect to the initial setup described by Moshinsky
\cite{Moshinsky52,Moshinsky76} aimed to study the matter-wave beam
dynamics, tends to wash out the oscillatory pattern of the beam
profile (An exception is the long-time revival of the diffraction
pattern described in \cite{DMM07}.) In this paper, we examine the
dynamics of a matter-wave beam in the presence of a moving mirror
(see Fig.~\ref{setup}) and identify characteristic regimes and
times, as well as quantum dynamical effects, such as the enhancement
of the self-modulation of the beam profile. With an infinite
velocity of the mirror, the usual diffraction in time result is
recovered.

In section \ref{sectionDIT}, we review briefly the Moshinsky shutter
problem and fix the notation; section \ref{sectionv} describes the
diffraction in time for mirrors moving at finite velocity. The paper
concludes with possible applications.

\section{Diffraction in time}\label{sectionDIT}

We shall first consider a quasi-monochromatic
atomic beam of momentum $\hbar k$ (or velocity $v_k=\hbar k/m$)
impinging on a totally reflecting shutter
at the origin $x=0$,
\beqa
\label{sine}
\psi_k(x',t'=0)=2i\sin(kx')\Theta(-x'),
\eeqa
where $\Theta(x)$ is the Heaviside step function. Such state
represents a standing matter-wave whose time evolution under sudden
removal of the mirror at $t=0$ we shall denote by
$\psi_k^{(\infty)}(x,t)$. Here the superscript $(\infty)$ underlines
the fact that the sudden removal of the mirror is equivalent to
displacing it to the right with infinite velocity. This wave is
clearly not normalized, but it can be considered as an elementary
component of a semi-infinite wave packet. (Following
customary practice, we shall liberally speak of its square modulus
as a ``density'' even if it is dimensionless.) Since for $t>0$ the
dynamics is free, it can be obtained using the superposition
principle
\beq
\label{superposition}
\psi_k^{(\infty)}(x,t)=\int_{-\infty}^{\infty}dx'
\K_0(x,t|x,t'=0)\psi(x',t'=0)
\eeq
with the free propagator
\beqa
\K_0(x,t\vert x',t')=\bigg[\frac{m}{2\pi i\hbar
(t-t')}\bigg]^{\frac{1}{2}}e^{i\frac{m(x-x')^{2}}{2\hbar (t-t')}}.
\eeqa
In a seminal paper \cite{Moshinsky52}, Moshinsky  proved that
\beqa
\label{eqDIT}
\psi_k^{(\infty)}(x,t)=M(x,k,t)-M(x,-k,t), 
\eeqa
where, the so-called Moshinsky function \cite{Moshinsky52,Moshinsky76}
\beq
M(x,k,t):=\frac{e^{i\frac{mx^{2}}{2 \hbar t}}}{2}w(-z),\quad
z=\frac{1+i}{2}\sqrt{\frac{\hbar t}{m}}\left(k-\frac{mx}{\hbar t}\right),
\eeq
is related to the Faddeyeva function $w(z):=
e^{-z^{2}}{\rm{erfc}}(-i z)$ \cite{Faddeyeva,AS65}. On physical
grounds it is clear that each of the $M$ functions corresponds to a
freely time-evolved cut-off plane-wave. Such solution entails the
well-known diffraction in time phenomenon which consists in a set of
oscillations in the beam profile, in dramatic contrast with the
classical case which is simply described by $\Theta(v_k t-x)$ for
$x>0$. [In principle a more complex ``classical analog'' may be
established using the Wigner function and classical trajectories
with negative weights \cite{MSS93,MMS99}, but in this work the
``classical case'' refers always to the 
incident beam formed by classical particles with fixed velocity
$v_k$.]

The asymptotics of the Moshinsky functions for $|x-\hbar kt/m|\rightarrow\infty$ can be found
from those of $w(z)$ when $z\rightarrow\infty$ \cite{MB00}.
In the classical region (where the classical beam density is non-zero),
$x\leq v_kt$, one finds in terms of the Gamma function
$\Gamma(y)$ that
\beqa
M(x,k,t)\sim e^{ikx-i\frac{\hbar k^2 t}{2m}}
+\frac{e^{i\frac{mx^2}{2\hbar t}}}{2\pi i}
\sum_{n=0}^{\infty}\frac{\Gamma\left(n+\frac{1}{2}\right)}{z^{2n+1}},
\eeqa
whereas in the complementary region $x>v_kt$ only the series survives.
Therefore the classical front plays an essential role in the quantum dynamics.

It was pointed out in \cite{DM05} that the initial state given by Eq. (\ref{sine}),
an eigenstate of a totally reflecting mirror,
maximizes the diffraction in time pattern with respect to the different types
of hard-wall mirrors. Moreover, the problem can be reformulated
as a sudden turn-on of a matter-wave source of the form
$\sigma(x,t)\propto\delta(x)\Theta(t)$ (an inhomogeneous term,
in the time-dependent Schr\"odinger equation) \cite{MoshinskyRMF52}.

\section{Moving mirror with finite velocity}\label{sectionv}

In this section we study the diffraction in time phenomenon when the
reflecting wall, acting as a perfect mirror,
is  displaced to the right with finite velocity, $x_m=vt$,
for $t>0$. It is useful to examine first the effect of the moving mirror for
a classical beam of particles with incident velocity $v_k$.
Two different regimes are possible
depending on whether the velocity of the mirror $v$ is larger or smaller
than $v_k$:  If $v>v_k$ no particle could hit the mirror for $t>0$
so there is no effect of the moving mirror in the beam dynamics.
This case is therefore classically equivalent to the sudden removal of the mirror;
The second case, $v<v_k$, is instead characterized by the occurrence of reflection. The particles
reflected by the moving mirror have velocity $2v-v_k$ (since $v-v_k$
is their velocity in the reference frame moving with the mirror),
and their front (marked by the first particles reflected just after the mirror starts moving, at $t=0_+=\epsilon>0$, when $\epsilon\to 0$)
is at $x_+=(2v-v_k)t$. This critical point may move to the right or left depending
on the sign of $2v-v_k$. A second critical space-time point at $x_-=-v_k t$ corresponds to the advancement leftwards of the last particle reflected with the stationary mirror at $t=0_-$. To the left of $x_-$ the beam is composed by particles incoming and reflected with velocities
$\pm v_k$, so that the effect of the moving mirror has not arrived yet.
In the intermediate region $x_-<x<x_+$, only right-moving particles are
found with velocity $v_k$. Finally, for $x>x_+$ there are incident particles
and reflected ones with
velocities $v_k$ and $2v-v_k$ respectively.

\begin{figure}
\includegraphics[width=7.5cm,angle=0]{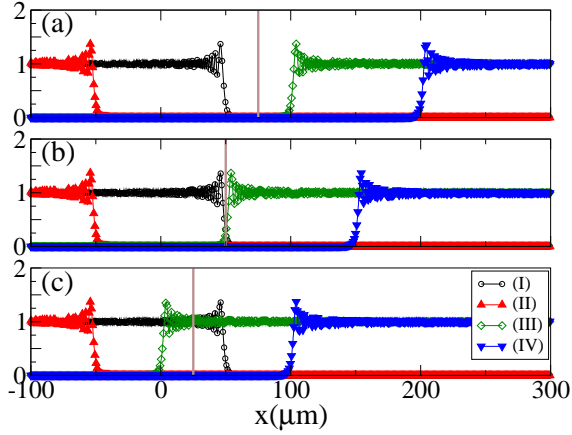}
  \caption{Density profile of the components involved in the quantum dynamics in the presence of a mirror.
The snapshots are taken  for a $^{87}$Rb atom beam moving at $v_k=1.0$ cm/s, impinging on a mirror which has
been displaced for $5$ ms at a) $v=1.5$, b) $1.0$, and c) $0.5$ cm/s. The position of the mirror is denoted
by a brown vertical line.
The components travel with $v_k$ (I) and $-v_k$ (II) as if in the absence of the mirror,
and the corresponding images that warrant the boundary condition at $x_m=vt$, travel with wavefronts at
$(2v-v_k)t$ (III) and $(2v+v_k)t$ (IV). Since (IV) moves into the forbidden region,
it soon stops playing a physical role.
}  \label{components}
\end{figure}
To cope with a moving mirror in quantum mechanics,
we take advantage of the result for the propagator
obtained with semiclassical arguments in \cite{LuzCheng92} and alternatively by summing
$\delta$-function perturbation series in path integrals \cite{Grosche93}, which turns out to be exact.
Its explicit form is
\beqa
\label{kv}
& & \K^{(v)}_{w}(x,t|x',t')=e^{-i\frac{m}{\hbar}\Big[v(x-vt)-v(x'-vt')
+\frac{v^2(t-t')}{2}\Big]}
\times\nonumber\\
& &\Big[\K_0(x-vt,t|x'-vt',t')
-\K_0(x-vt,t|-x'+vt',t')\Big].\nonumber\\
\eeqa
It is of the ``collapsed'' kind \cite{Crandall83}, since it is obtained from
just two classical paths.

Let us assume the Moshinsky initial condition (1) of a
quasi-monochromatic beam incident on the shutter, which is located
at the origin at time $t'=0$. At a later time the wavefunction can
be calculated using the integral equation (\ref{superposition}) with
the kernel of Eq.(\ref{kv}) and the initial state
$\psi(x',t'=0)=2i\sin(kx')\Theta(-x')$.

Performing the integrals, one finds
\beqa
& & \psi_{k}^{(v)}(x,t)=
e^{i\frac{mvx}{\hbar}-i\frac{mv^2t}{2\hbar}}
\nonumber\\
& & \times\bigg[
M\left(x-vt,k-\frac{mv}{\hbar},t\right)
-M\left(x-vt,-k-\frac{mv}{\hbar},t\right)\nonumber\\
& &-M\left(vt-x,k-\frac{mv}{\hbar},t\right)
+M\left(vt-x,-k-\frac{mv}{\hbar},t\right)\bigg],
\nonumber\\
\label{vmirror}
\eeqa
for the physical region $x\le x_m$.
The physical wave function is of course zero in the forbidden region
but it is
useful to consider formally Eq. (\ref{vmirror}) also at $x>x_m$
for the simple analysis of ``image'' points and term contributions.
The first two terms, $M_I$ and $M_{II}$ for short,
describe  the free time evolution of
the beam in the absence of the mirror,
$\psi^{(\infty)}(x,t)$, with wavefronts at $x=\pm v_k t$ (note that $v_k$t is
in the forbidden region when $v_k>v$);
whereas the third and fourth terms, $M_{III}$ and $M_{IV}$ for short,
are their
corresponding images with respect
to the position of the mirror $x_m$, and are relevant for $x\gtrsim(2v\pm v_k)t$.
The corresponding ``densities'' are shown in Fig.~\ref{components}.
Clearly $M_{IV}$, whose wavefront is at $(2v+v_k)t$ represents in
general a minor contribution for it travels into the forbidden
region beyond $x_m$ for all $v$, whereas the front of $M_{III}$, at
$x_+=(2v-v_k)t$ enters into the forbidden region for $v>v_k$.

In a reference frame moving with the mirror,  the dynamics arises as a result of
a kick imparted on the beam.
%
\begin{figure}
\includegraphics[width=7.5cm,angle=0]{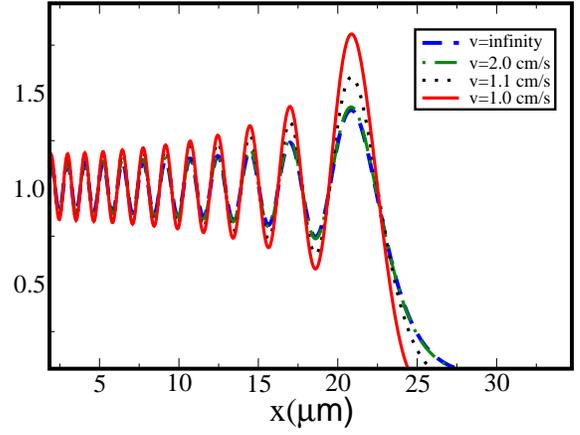}
  \caption{
Characteristic modulation of the density profile of a matter-wave
beam of $^{87}$Rb atoms induced by a totally reflecting mirror which
is displaced at constant velocity $v>v_k$ ($v_k=1.0$ cm/s). The
oscillatory pattern is maximized as $v$ assumes the velocity of the
beam. The case $v\rightarrow\infty$ reduces to the diffraction in
time setup with the lowest fringe visibility. } \label{enhancement}
\end{figure}
%
Let us check first the consistency of the mathematical result
in the limits of very slow and very fast walls. For a very slow wall,
$v\sim0$, the result reduces to that of a
fixed wall, with $\psi_{k}^{(v)}(x,t)\sim
\psi_k(x,t)$,
as it follows using the exact relation
\beq
\label{planewave}
M(x,k,t)+M(-x,-k,t)=e^{ikx-i\frac{\hbar k^2t}{2m}}.
\eeq
($M_I+M_{IV}$ give an incident plane wave and $M_{II}+M_{III}$ the
reflected wave with a global minus sign to produce the sine in Eq.
(\ref{sine}).) In the opposite case where the velocity of the moving
wall is much bigger than that of the incident beam the two last
terms in Eq. (\ref{vmirror}) become negligible and one has
$\psi_{k}^{(v)}(x,t)\sim M(x,k,t)
-M(x,-k,t)=\psi_{k}^{(\infty)}(x,t)$. This is the limit of
infinitely fast removal of the shutter, which lead to the discovery
of diffraction in time.

As for the classical beams, two different velocity regimes can be
clearly distinguished; as expected, in both velocity regimes the
quantum wave shows characteristic diffraction in time patterns. When
the velocity of the mirror is larger than the mean velocity of the
beam nothing happens classically different from $v=\infty$ because
of the absence of reflection, but quantally the amplitude of the
oscillations increases whereas its spacing is unaffected. Figure
\ref{enhancement} shows that the presence of the mirror leads to an
enhancement of the diffraction in time as the velocity of the wall
approaches the mean velocity of the beam. The maximum intensity
$P_{max}$ of the main peak increases as $v\rightarrow v_k$ reaching
an upper bound which is $1.816$ times the stationary value, whereas
in the Moshinsky solution, Eq.(\ref{eqDIT}), the increment is
limited to a $1.37$ times the stationary value \cite{note}.
Therefore the dominant effect is the enhancement of diffraction in
time pattern. This can be understood as follows: As $v\rightarrow
v_k$ and for $0<x<x_m$ we can neglect the second and fourth terms in
Eq. (\ref{vmirror}), \beqa & & \psi_{k}^{(v)}(x,t)\simeq
e^{i\frac{mvx}{\hbar}-i\frac{mv^2t}{2\hbar}}
\nonumber\\
& & \times\big[
M\left(x-vt,0,t\right)
-M\left(vt-x,0,t\right)\big].
\eeqa
Using Eq. (\ref{planewave}) together with $M(x,k,t)=e^{ikx-i\frac{\hbar k^2t}{2m}}{\rm erfc}
[\frac{1+i}{2}\sqrt{\hbar t/m}(k-mx/\hbar t)]$,
and ${\rm erfc}(z)+{\rm erfc}(z)=1$, this may be written as
\beqa
& & \psi_{k}^{(v)}(x,t)\simeq
e^{i\frac{2mvx}{\hbar}-i\frac{mv^2t}{\hbar}}
{\rm erfc}\bigg[\frac{1+i}{2}\sqrt{\frac{\hbar t}{m}}\left(\frac{mv}{\hbar}
-\frac{mx}{\hbar t}\right)\bigg].\nonumber\\
\eeqa
%
\begin{figure}
\includegraphics[width=7.5cm,angle=0]{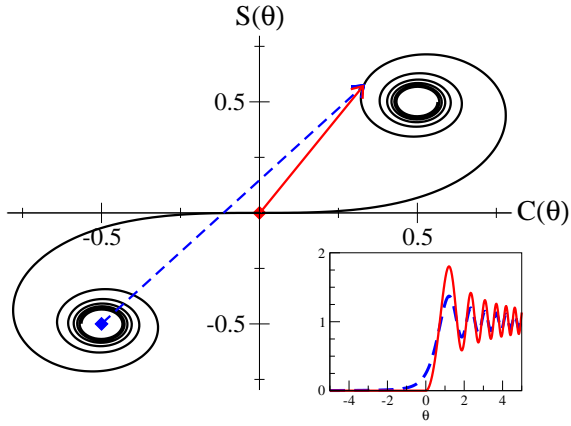}
  \caption{Universal representation of the beam profile
for ordinary (dashed line) and enhanced (solid line) diffraction
in time in terms of the Cornu spiral, when the velocity of the mirror approaches that of the beam.
The inset shows the probability density
of both beam profiles as a function of the variable $\theta$.
}  \label{cornu}
\end{figure}
%
%
Its absolute square value admits a simple geometric interpretation
in terms of the Cornu spiral or clothoid, which is the curve that
results from a parametric representation of the Fresnel integrals,
$S(\theta)$ versus $C(\theta)$ as shown in Fig.~\ref{cornu}.

Introducing $\theta=\sqrt{\hbar t/m\pi}(\hbar kt/m-x)$,
the universal representation of the beam profile reads
\beqa
|\psi_{k}^{(v)}(x,t)|^2\simeq
2\big\{[S(\theta)]^2+[C(\theta)]^2\big\},
\eeqa
this is, twice the distance from the origin (where $\theta=0$, at
the position of the mirror) to any point of the spiral with
$\theta>0$ (first quadrant), being zero elsewhere \cite{erfc}. Moshinsky
\cite{Moshinsky52} showed that whenever the $-k$ component in the
ordinary diffraction in time problem, described
by $\psi_{k}^{(\infty)}(x,t)$, can be neglected, the beam profile
admits also a universal representation in the form,
\beqa
|\psi_{k}^{(\infty)}(x,t)|^2\simeq
\frac{1}{2}\bigg\{\left[S(\theta)+\frac{1}{2}\right]^2+\left[C(\theta)+\frac{1}{2}\right]^2\bigg\},
\eeqa
see Fig.\ref{cornu}, which is half the distance from the point $(-1/2,-1/2)$ to the Cornu
spiral for arbitrary $\theta$.
%
\begin{figure}
\includegraphics[width=7.5cm,angle=0]{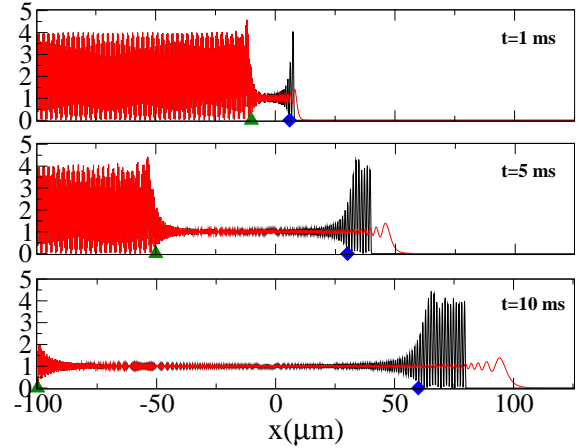}
  \caption{Transients in the density profile $|\psi_v(x,t)|^2$
  of an atom beam when the velocity
of the mirror is less than the mean of the beam (solid line)
versus the sudden-removal case (dashed line),
with $t=10$ ms, $v=0.8$ cm/s and $v_k=1.0$ cm/s.
The triangles point out the classical position of a particle moving with $-v_k$,
and the diamonds the location of a particle after
bouncing off from the mirror, $(2v-v_k)t$.
The oscillations of the matter-wave front reach a visibility $V=1$
tending to the asymptotic form.
}  \label{dyn}
\end{figure}
%
It follows that the frequency of the oscillations is the same
for the ordinary and enhanced diffraction in time in the limit $v\rightarrow v_{k}^+$.
Moreover the width of the fringes $\delta x$ is also common to both cases, and can be estimated
from the intersection between the classical and quantum probability densities \cite{Moshinsky52,DM05},
leading to a dependence of the form
\beqa
\delta x\propto(\pi\hbar t/m)^{1/2}.
\eeqa
Note that the intensity of the beam tends to unity in both cases for $\theta\rightarrow\infty$ (i.e., away from the mirror).

From a more physical perspective the enhancement is due to the
contribution of the image $M(vt-x,0,t)$ of the main term
$M(x-vt,0,t)$, i.e., to a quantum reflection contribution (see Fig.
\ref{components}b). In the classical limit, the trajectory closest
to the mirror remains unaffected by the presence of the mirror if
$v\ge v_k$; however in quantum mechanics the front of $M(x-vt,0,t)$
is not sharply localized and in addition some reflection occurs.
This effect becomes smaller when $v-v_k$ increases because of the
displacement of the image term front into the forbidden region;
equivalently, the front tail of the leading Moshinsky function lags
behind the mirror. As a result no reflection will occur, as
demonstrated in Fig. \ref{components}a.

In the opposite regime, if $v_k>v$, the
wavefront at $x_+$ bounces off from the mirror leading to an
interference enhancement and the progressive construction of the new
scattering state with velocities $v_k$ (incident) and $2v-v_k$
(reflected) in the domain $x_+<x<vt$.
The dynamical evolution is illustrated in Figure \ref{dyn}:  as
a consequence of the reflection, the interference pattern can reach
values up to four times the stationary one. Indeed, during the
evolution of the beam three distinct regions can be identified,
$|\psi_{k}^{(v)}(x,t)|^2\sim$
\beqa
\left\{\begin{array}{l l l}
4\sin(kx) &,\;\; x\leq x_-\\
1 &,\;\; x_- <x\leq x_+\\
4\sin[(k-v)(x-vt)] &,\;\; x_+<x<vt
\end{array}\right.,
\eeqa
where,
the separating limits are not sharp
but described by the corresponding Moshinsky functions.
%
\begin{figure}
\includegraphics[width=7.5cm,angle=0]{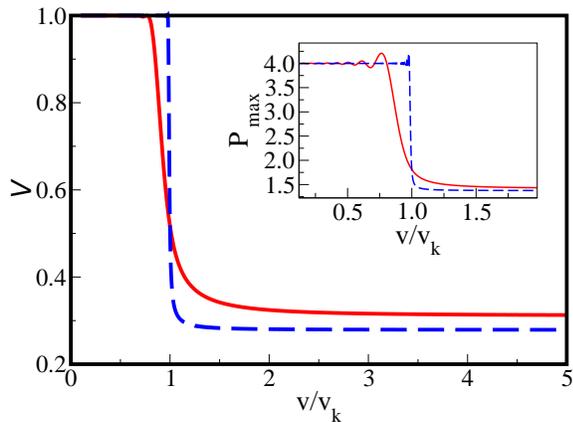}
  \caption{Visibility of the fringes in the density profile of an atom beam for $v_k=
0.1$ (solid line), and $1.0$ cm/s (dashed line) at time $t=100$ ms.
The inset shows the maximum of the main fringe for the same range of $v/v_k$ values.
}  \label{visibility}
\end{figure}
%

To characterize the amplitude of the oscillations we may consider
the visibility of the main fringe, defined by
\beqa
V=\frac{P_{max}-P_{min}}{P_{max}+P_{min}},
\eeqa
where $P_{min}$ is the first minimum of the wave density behind the
matter-wave front. Figure \ref{visibility} shows how for a fixed time, 
this measure exhibits a monotonic decay with increasing ratio $v/v_k$, 
reaching the sudden-removal result as
$v/v_k\rightarrow\infty$.

\section{Conclusion and Discussion}

A standing matter wave suddenly released, develops during its
propagation an oscillatory pattern in the density profile, a
phenomenon known as diffraction in time. We have shown that if the beam 
is released by moving the mirror at finite velocity 
there is an enhancement of such effect 
and of the visibility of the corresponding fringes. 
The result is relevant for recently proposed atom beam techniques,
such as beam slowing with mirrors on spinning rotors \cite{Raizen06}
or conveyor belts \cite{Guery06}, and schemes for atom
interferometry in time domain when operating with ultracold
velocities. Being intrinsically a matter-wave effect, the
enhancement of the diffraction in time could be observed with
ultracold neutrons as well \cite{ucn}. 

Further extension of this work can be envisaged 
to deal with the dynamics of finite pulses \cite{Moshinsky76,DM05,DMM07}, 
and the use of time dependent external fields for controlled 
transport of matter waves \cite{DM06b,Song03}.

%

\begin{acknowledgments}
This work has been supported by 
DGES-Spain (FIS2006-10268-C03-01) and UPV-EHU (grant 00039.310-15968/2004).
AC acknowledges discussions with M. G. Raizen and S. Godoy,
the hospitality of M. Kleber and H. Friedrich's group at
the Physik-Department of the Technische Universit{\"a}t M{\"u}nchen
where part of this work was carried out,
and a fellowship from the Basque Government (BFI04.479)
\end{acknowledgments}


\begin{thebibliography}{57}
\bibitem{note} We have also considered the case of a semi-transparent mirror of the form
$V(x)=\eta\delta(x-vt)$. The resulting dynamics is still exactly solvable,
and shows that the enhancement of the diffraction is maximized in the limit
$\eta\rightarrow\infty$, discussed here.
\bibitem{erfc} We use the relation $C(z)+iS(z)=\frac{1+i}{2}[1-{\rm erfc}(\sqrt{\pi}(1-i)z/2)]$ \cite{AS65}.
\end{thebibliography}
\end{document}